\documentclass[preprint]{jpsj2} 

\newcommand{\Prob}{\mbox{Prob}}
\newcommand{\sgn}{\mbox{sgn}}
\newcommand{\erf}{\mbox{erf}}
\renewcommand{\vec}[1]{\mbox{\boldmath $#1$}}

\title{Theory of Recurrent Neural Network with Common Synaptic Inputs}

\author{Masaki \textsc{Kawamura}$^{1}$
\thanks{E-mail: kawamura@sci.yamaguchi-u.ac.jp}
Michiko \textsc{Yamana}$^{2}$
and
Masato \textsc{Okada}$^{3,4,5}$
}

\inst{$^{1}$Faculty of Science, Yamaguchi University, 1677-1 Yoshida, Yamaguchi 753-8512 \\
$^{2}$ System Engineering Research Laboratory, Central Research 
Institute of Electric Power Industry, 
2-11-1 Iwadokita, Komae, Tokyo 201--8511 \\
$^{3}$ Department of Complexity Science and Engineering,
Graduate School of Frontier Sciences,
The University of Tokyo,
5-1-5 Kashiwanoha, Kashiwa, Chiba 277-8562 \\
$^{4}$ Laboratory for Mathematical Neuroscience, RIKEN Brain Science
Institute, 2-1 Hirosawa, Wako, Saitama 351-0198 \\
$^{5}$ "Intelligent Cooperation and Control," PRESTO, JST
}

\abst{ We discuss the effects of common synaptic inputs in a recurrent
neural network.  Because of the effects of these common synaptic inputs,
the correlation between neural inputs cannot be ignored, and thus the
network exhibits sample dependence. 
Networks of this type do not have well-defined thermodynamic limits, and 
self-averaging breaks down.
We therefore need to develop a suitable theory
without relying on these common properties.
While the effects of the common synaptic inputs have been analyzed in
layered neural networks, it was apparently difficult to analyze these 
effects in recurrent neural networks due to feedback connections. 
We investigated 
a sequential associative memory model as an example of recurrent
networks and succeeded in deriving a macroscopic dynamical description
as a recurrence relation form of a probability density function.  }

\kword{common synaptic inputs, recurrent neural networks, 
probability density function, correlated firings, sample dependence}

\begin{document}
\maketitle

\section{Introduction}

Synfire chains, namely, synchronous firings of neurons, can be observed
in the brain \cite{Abeles1991}. Diesmann {\itshape et
al.}\cite{Diesmann_etal1999} and C\^ateau and Fukai
\cite{CateauFukai2001} discussed conditions for propagating the
synchronous firings between layers in layered neural networks, while
Amari {\itshape et al.} considered common synaptic inputs to neurons in
the layered neural networks and discussed correlated firings of neurons
\cite{Amari_etal2003}.
These studies are based on theoretical models, and the biological
structure of the synfire chains or the common synaptic inputs remain to
be elucidated. In order to understand the structure, theoretical models
must be analyzed. We therefore discuss the effects of the
common synaptic inputs on an associative memory model from a theoretical
viewpoint. In order to analyze these effects, the structure of our model
is simple, unlike that for synfire chain models.

Using the common synaptic inputs, the sums of inputs to neurons are
correlated.  The firings of the neurons are, therefore, also correlated,
and with an infinite number of neurons there is no thermodynamic limit
and sample dependence appears \cite{Amari_etal2003,Sakai2002}.  All
solvable models, including neural networks, that have been discussed in
the statistical mechanics literature have been analyzed by applying the
independence of units or neurons at the thermodynamic limit.  There are
few theoretical approaches, however, that address sample dependence.
Yamana and Okada \cite{YamanaOkada2005} introduced uniform common
synaptic inputs that depend on preneurons to the layered associative
memory model, and was able to derive the probability density function (PDF)
for its macroscopic states.  This PDF allows for the analysis of
dynamics with sample dependence in the layered associative memory model.

In the layered associative memory model, since synaptic connections
within a layer are independent of each other, no correlation occurs
between common synaptic inputs on different layers. However, in 
recurrent neural networks that contain feedback connections, correlations
between common synaptic inputs at different times cannot be
ignored. Theoretical analysis in such cases might be rendered difficult,
and in fact it is indeed hard to analyze qualitatively the effect of
common synaptic inputs in an autoassociative memory model
\cite{YoshiokaShiino1997}.

In recurrent neural networks, correlated connections can also be
found in asymmetric synaptic connections
\cite{Peretto1988,YoshiokaShiino1997}, e.g.,
\begin{equation}
 J_{ij}\propto \sum_{\mu}\left(\xi_i^{\mu}+a\right)\left(\xi_j^{\mu}+b\right)
  =\sum_{\mu}\xi_i^{\mu}\xi_j^{\mu}+a\xi_j^{\mu}+b\xi_i^{\mu}+ab,
\end{equation}
where $\vec{\xi}^{\mu}=(\xi^{\mu}_1,\cdots,\xi^{\mu}_N)^T$ represents
the $\mu$th memory pattern. The terms $a\xi_j^{\mu}$ and $b\xi_i^{\mu}$
indicate the connections that depend on pre- and postneurons,
respectively. These terms can be considered to be noise to neurons
\cite{YoshiokaShiino1997}. Since the preneuron-dependent connection
leads to correlated firings of neurons, we take particular note of the
term $a\xi_j^{\mu}$. Moreover, we reduce this term to one that is
independent of the memory patterns, $w_j$. 

In this paper, we discuss a sequential associative memory model that is
also a recurrent neural network
\cite{Amari1988b,During1998,KitanoAoyagi1998,Katayama2001,Kawamura2002}.
The associative memory model stores memory patterns in the synaptic
connections; that is, the synaptic connections are not uniform, but they
do have a structure.  Moreover, the synaptic connections are time
invariant, unlike those for layered networks. We found, however, that
time correlations of states in this model can be ignored when a memory
pattern is retrieved, since the model retrieves a different pattern
sequentially each time. The common synaptic inputs at different times
can, therefore, be assumed to be independent. Under this consideration,
we have succeeded in deriving a recurrence relation form of the
probability density function for macroscopic states in the sequential
associative memory model.

\section{Sequential Associative Memory Model}

Consider a sequential associative memory model \cite{Amari1988b}
consisting of $N$ units or neurons. The state of the units takes
$x^t_i=\pm1$ and is updated synchronously by
\begin{equation}
 x_i^{t+1} = F\left(\sum_{j=1}^NJ_{ij}x_j^t\right),
  \label{eqn:dynamics}  
\end{equation}
where the output function is $F(h)=\sgn(h)$, and  $J_{ij}$ is a synaptic
connection from the $j$th neuron to the $i$th neuron, and given by
\begin{equation}
 J_{ij}=\frac1N\sum_{\mu=0}^{p-1}\xi^{\mu+1}_i\xi^{\mu}_j +w_j,
  \label{eqn:Jij}
\end{equation}
where $\vec{\xi}^{p}=\vec{\xi}^0$. The first term on the rhs represents
the coupling as in the existing sequential associative memory model
\cite{Amari1988b,During1998,KitanoAoyagi1998,Katayama2001,Kawamura2002}.
It stores $p$ random patterns
$\vec{\xi}^{\mu}=(\xi^{\mu}_1,\cdots,\xi^{\mu}_N)^T$ so as to retrieve
the patterns as
$\vec{\xi}^0\to\vec{\xi}^1\to\cdots\vec{\xi}^{p-1}\to\vec{\xi}^0$
sequentially. The second term on the rhs, $w_j$, 
represents preneuron-dependent coupling.
From eqs. (\ref{eqn:dynamics}) and (\ref{eqn:Jij}), we obtain
\begin{eqnarray}
 x_i^{t+1} &=& F\! \left(\frac1N\sum_{\mu=0}^{p-1}\sum_{j=1}^N
                 \xi^{\mu+1}_i\xi^{\mu}_jx_j^t+
                 \sum_{j=1}^Nw_jx_j^t\right) .
\label{eqn:x_xi}
\end{eqnarray}
Let the second term on the rhs be
\begin{equation}
 \eta_{t}=\sum_{j=1}^Nw_jx_j^t .
  \label{eqn:eta}
\end{equation}
We call $\eta_{t}$ the {\itshape common synaptic input}, since it is
independent of index $i$ and affects all neurons equally.
Of course one can consider $\eta_{t}$ to be an external input coming
from the outside system, which might be independent of preneurons
$x_j^{t}$. 
In order to analyze the dynamics theoretically, 
we will assume that the coupling $w_j$ obeys the Gaussian
distribution with ${\cal N}(0,\delta^2/N)$.
Therefore, the common synaptic inputs act like noise in this case.

The number of neurons is given by $p=\alpha N$. We call $\alpha$ the
{\itshape loading rate}. Each component of the memory patterns is
assumed to be an independent random variable that takes a value of
either $+1$ or $-1$ according to the probability
\begin{equation}
 \Prob\left[\xi_i^{\mu}=\pm1\right]=\frac{1}{2} .
\end{equation}
We define the overlap by the direction cosine between the state
$\vec{x}^t$ and the retrieval pattern $\vec{\xi}^t$ at time $t$,
\begin{equation}
 m_t = \frac{1}{N}\sum_{i=1}^N \xi_i^t x_i^t ,
  \label{eqn:mt}
\end{equation}
and determine the initial state $\vec{x}^0$ according to the probability
distribution
\begin{equation}
 \Prob[x^0_i=\pm1] = \frac{1\pm m_0 \xi^0_i}{2}.
\end{equation}
Therefore, the overlap between the pattern $\vec{\xi}^0$ and the initial
state $\vec{x}^0$ is $m_0$.  The network state $\vec{x}^t$ at time $t$
is expected to be near the pattern $\vec{\xi}^t$ when the initial
overlap $m_0$ is large and the loading rate is under its storage
capacity.

\section{Theory}

\subsection{Macroscopic state}

Let us derive the macroscopic state equations in the case of 
dynamics with sample dependence. 
From eqs. (\ref{eqn:x_xi})--(\ref{eqn:mt}), we obtain
\begin{eqnarray}
 x^{t+1}_i &=& F\left(\xi^{t+1}_im_t+z^t_i+\eta_t\right) , \\
 z^t_i &=& \frac{1}{N}\sum_{\mu\neq t}\sum_{j=1}^{N}
  \xi^{\mu+1}_i\xi^{\mu}_j x^t_j ,
\end{eqnarray}
where $z^t_i$ is a crosstalk noise term. We assume that the crosstalk
noise obeys the Gaussian distribution with mean 0 and variance
$\sigma_t^2$ according to the statistical neurodynamics
\cite{Amari1988,Amari1988b,Okada1995,Okada1996}.  The state $x^{t+1}_j$
is expanded in terms of $\xi^{\mu}_j$ as follows:
\begin{eqnarray}
 z^{t+1}_i &=& \!\! \frac{1}{N} \!\! \sum_{\mu\neq t+1}\sum_{j=1}^{N}
  \xi^{\mu+1}_i\xi^{\mu}_j F\left(\xi^{t+1}_jm_t+z^t_j+\eta_t\right) ,
 \\
 &=& \frac{1}{N} \sum_{\mu\neq t+1}\sum_{j=1}^{N}
  \xi^{\mu+1}_i\xi^{\mu}_j x^{t+1,(\mu)}_j 
  +\frac{1}{N} \sum_{\mu\neq t+1}\sum_{k=1}^{N}
  \xi^{\mu+1}_i\xi^{\mu-1}_kx^t_kU_{t+1}, 
\end{eqnarray}
where 
\begin{eqnarray}
 x^{t+1,(\mu)}_j &=& F\left(\frac{1}{N}\sum_{\nu\neq \mu}\sum_{k=1}^{N}
                       \xi^{\nu+1}_j\xi^{\nu}_k x^t_k+\eta_t\right) , \\
 U_{t+1} &=&  \frac{1}{N}\sum_{j=1}^N
  F'\left(\xi^{t+1}_jm_t+z^t_j+\eta_t\right) .
\end{eqnarray}
Therefore, the variance of the crosstalk noise becomes 
\cite{KitanoAoyagi1998,Kawamura2002}
\begin{equation}
 \sigma_{t+1}^2 = E\left[\left(z^{t+1}_i\right)^2\right] = \alpha+U_{t+1}^2\sigma_{t}^2 .
\end{equation}

Let us next consider the initial state to be $\vec{x}^0=\vec{\xi}^0$. In
this case, since the pattern $\vec{\xi}^t$ is retrieved at time $t$ and
the memory patterns are independent of each other, we can assume that
$\vec{x}^t \approx\vec{\xi}^t$ and the state $x^t_i$ can become
independent of each other. When the correlation between $w_j$ and $x^t_j$
can be neglected, the common synaptic inputs $\eta_t$ are independent
with respect to time $t$. Therefore, $\eta_t$ are {\itshape iid} and
they obey the Gaussian distribution with ${\cal N}(0,\delta^2)$.  First
of all, we will discuss the case in which $\eta_t$ is given. Here, 
$m_{t+1}$ and $\sigma_{t+1}$ can be represented as functions of
$m_t$, $\sigma_t$ and  $\eta_t$ :
\begin{eqnarray}
 m_{t+1}(m_t,\sigma_t,\eta_t) &=& \int \!\! D_{z} 
  \left<\xi^{t+1} F\left(\xi^{t+1}m_t+\sigma_t z+\eta_t\right)\right>_{\xi}, 
  \\
 &=& \frac{1}{2} \left[\erf\left(u\right)+\erf\left(v\right)\right],
  \label{eqn:merf} \\
 U_{t+1}(m_t,\sigma_t,\eta_t) &=& \int D_{z}
  \left<F'\left(\xi^{t+1}m_t+\sigma_t z+\eta_t\right)\right>_{\xi} ,
  \nonumber \\
 &=& \!\!\! \frac{1}{\sqrt{2\pi}\sigma_t} 
  \left[\exp\left(-u^2\right)+\exp\left(-v^2\right)\right] ,
  \\
 \sigma_{t+1}^2(m_t,\sigma_t,\eta_t) &=&
  \alpha + U_{t+1}^2(m_t,\sigma_t,\eta_t)\sigma_t^2(m_{t-1},\sigma_{t-1},\eta_{t-1}), 
  \label{eqn:sigma_t} 
\end{eqnarray}
where 
$D_{z}=\frac{dz}{\sqrt{2\pi}} \exp\left(-\frac{z^2}{2}\right)$ and 
$u=\left(m_t+\eta_t\right)/\sqrt{2}\sigma_t$,
$v=\left(m_t-\eta_t\right)/\sqrt{2}\sigma_t$; 
$\left<\cdot\right>_{\xi}$ denotes the average over $\xi$.

\subsection{Probability density function}
\label{sec:pdf}

From eqs.~(\ref{eqn:merf})--(\ref{eqn:sigma_t}) we can evaluate the
dynamics for various values of $\delta$. In the case of $\delta=0$, the
behavior is deterministic as with the existing sequential associative
memory model. In the case of $\delta>0$, the behavior changes
drastically. Furthermore, $m_t$ and $\sigma_t$ are distributed, and these
distributions are described as the probability density function
$p\left(m_t,\sigma_t,\eta_t\right)$.  As described above, since $m_t$
and $\sigma_t$ are independent of $\eta_t$, 
the probability density function is decoupled as
\begin{equation}
p\left(m_t,\sigma_t,\eta_t\right)
 =p\left(m_t,\sigma_t\right)p\left(\eta_t\right). 
\end{equation}
We can, therefore, obtain the PDF by
\begin{eqnarray}
 p\left(m_{t+1},\sigma_{t+1}\right) &=& \!\!\!
  \int dm_t d\sigma_t d\eta_t p\left(m_t,\sigma_t\right)
  p\left(\eta_t\right) \nonumber \\
 &\times& \delta\left(m_{t+1}-m_{t+1}(m_t,\sigma_t,\eta_t)\right) 
 \delta\left(\sigma_{t+1}-\sigma_{t+1}(m_t,\sigma_t,\eta_t)\right) ,
  \label{eqn:Pnext}
\end{eqnarray}
where $\delta(\cdot)$ denotes the Dirac's delta function.
The PDF of $p\left(\eta_t\right)$ is given by 
\begin{equation}
 p\left(\eta_t\right) = \frac{1}{\sqrt{2\pi}\delta}
  \exp\left(-\frac{\eta_t^2}{2\delta^2}\right) .
\end{equation}
We combine the terms of $\eta_t$ into kernel function
$K\left(m_{t+1},\sigma_{t+1};m_t,\sigma_t\right)$: 
\begin{eqnarray}
 p\left(m_{t+1},\sigma_{t+1}\right)
 &=&  \!\! \int \!\! dm_t d\sigma_t p\left(m_t,\sigma_t\right) 
  K\left(m_{t+1},\sigma_{t+1};m_t,\sigma_t\right) \! ,
 \label{eqn:PnextK} \\
 K\left(m_{t+1},\sigma_{t+1};m_t,\sigma_t\right)
 &=& \!\! \int \!\! d\eta_tp\left(\eta_t\right)
  \delta\left(m_{t+1}-m_{t+1}(m_t,\sigma_t,\eta_t)\right) \nonumber \\
 && \times \delta\left(\sigma_{t+1}-\sigma_{t+1}(m_t,\sigma_t,\eta_t)\right) .
\end{eqnarray}
The kernel function $K\left(m_{t+1},\sigma_{t+1};m_t,\sigma_t\right)$
can be evaluated analytically. Let $m_t^*, \sigma_t^*$ be the value of 
$m_t, \sigma_t$ satisfying eqs.~(\ref{eqn:merf})--(\ref{eqn:sigma_t}).
Then, $K\left(m_{t+1},\sigma_{t+1};m_t,\sigma_t\right)$ becomes
\begin{eqnarray}
 K\left(m_{t+1},\sigma_{t+1};m_t,\sigma_t\right)
  = \int d\eta_tp\left(\eta_t\right) \frac{2\pi\eta_t^2
  \sqrt{2\pi\alpha+\left(e^{-u^2}+e^{-v^2}\right)^2}}
  {\left(u-v\right)^2\left(e^{-u^2}+e^{-v^2}\right)e^{-u^2-v^2}},
\end{eqnarray}
where 
$u=\left(m_t^*+\eta_t\right)/\sqrt{2}\sigma_t^*$ and
$v=\left(m_t^*-\eta_t\right)/\sqrt{2}\sigma_t^*$.
Our PDF agrees with the PDF for the layered associative memory model
obtained by Yamana and Okada \cite{YamanaOkada2005}. 

\section{Effect of Common Synaptic Inputs}

We demonstrate the effect of the common synaptic inputs in our model
with computer simulations. The effect of the inputs depends on the
variance $\delta^2$. In the case of $\delta=0$, the dynamical behaviors
are uniquely determined according to the initial states. On the other
hand, in the case of $\delta>0$, since there exists the correlation
between the inputs to neurons, sample dependence arises. That is, the
dynamical behaviors are not determined according to the initial states,
and the model either succeeds or fails to retrieve the memory pattern
from the same initial state.

\begin{figure}[bt]
 \begin{center}
  \includegraphics[width=75mm]{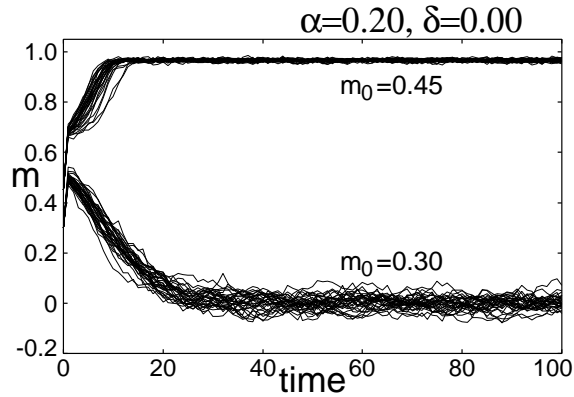} 
 \end{center}
 \caption{Time evolutions of overlap without common synaptic inputs
 ($\delta=0$) for $m_0=0.30$ and $0.45$, where $\alpha=0.20$ and $N=5,000$.}
 \label{fig:delta0}
\end{figure}

First, we show the time evolutions of overlap when there is no common
synaptic input ($\delta=0$). Figure~\ref{fig:delta0} shows $30$ samples of
overlap $m_t$ for initial overlaps $m_0=0.30$ and $0.45$, where the
loading rate is $\alpha=0.20$ and the number of neurons is $N=5,000$.
While there are fluctuations in the overlaps, they are caused by a
finite number of neurons; the larger the number of neurons is, the
smaller the fluctuations are. Therefore, no sample dependence can be
seen in the case of $\delta=0$.

\begin{figure}[tb]
 \begin{center}
  \includegraphics[width=75mm]{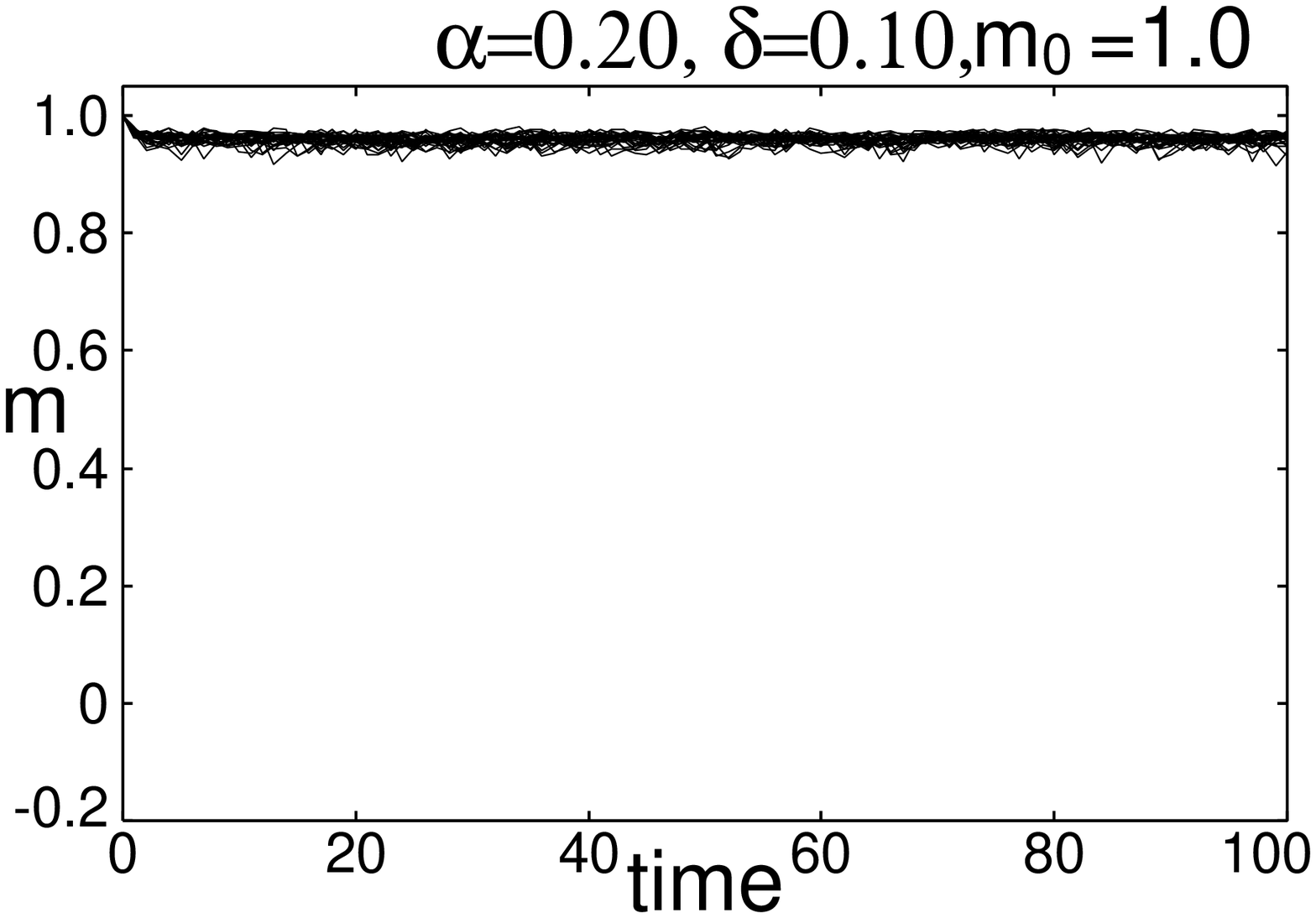} \\
  (a) $\delta=0.1$
  
  \includegraphics[width=75mm]{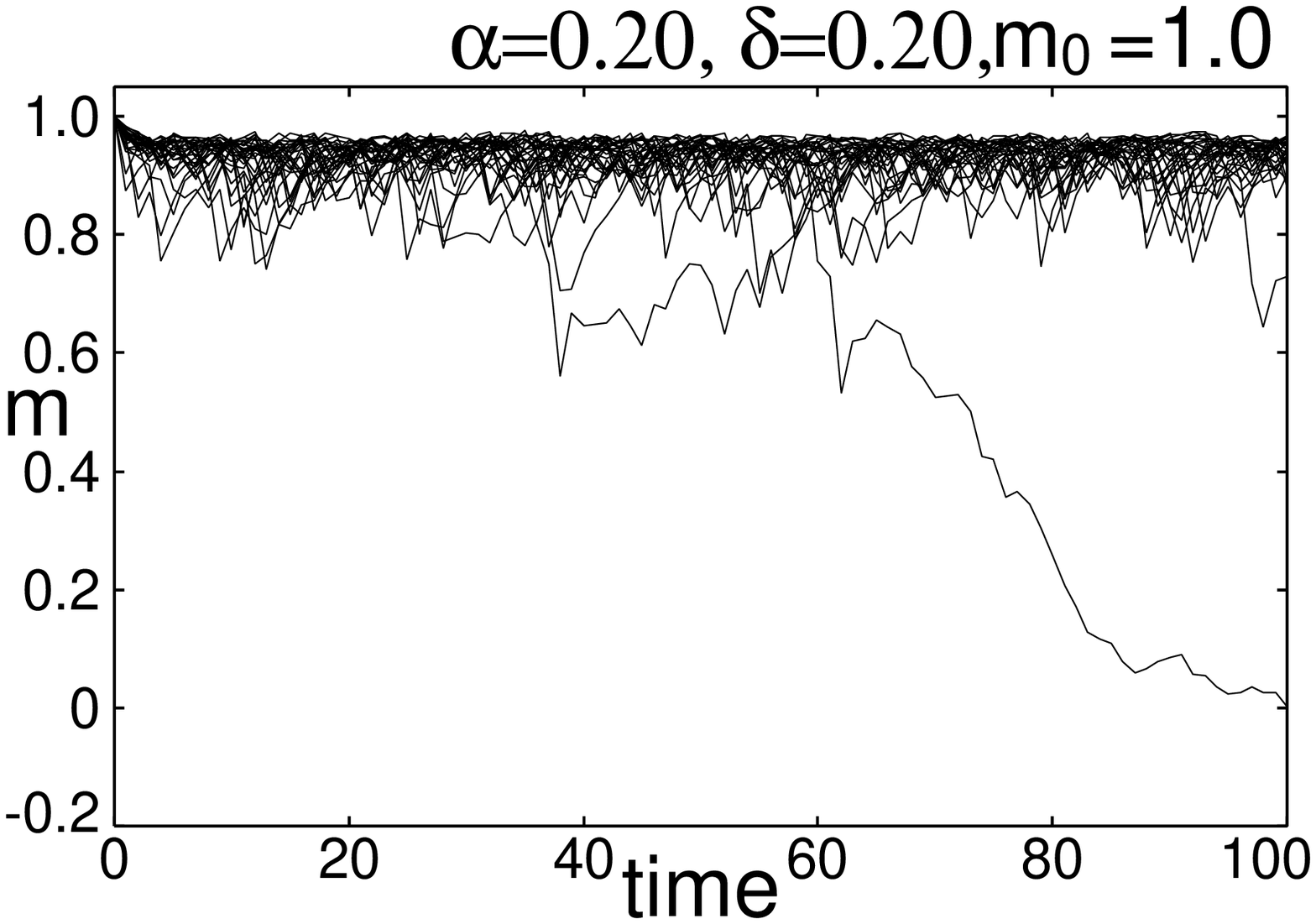}\\
  (b) $\delta=0.2$

  \includegraphics[width=75mm]{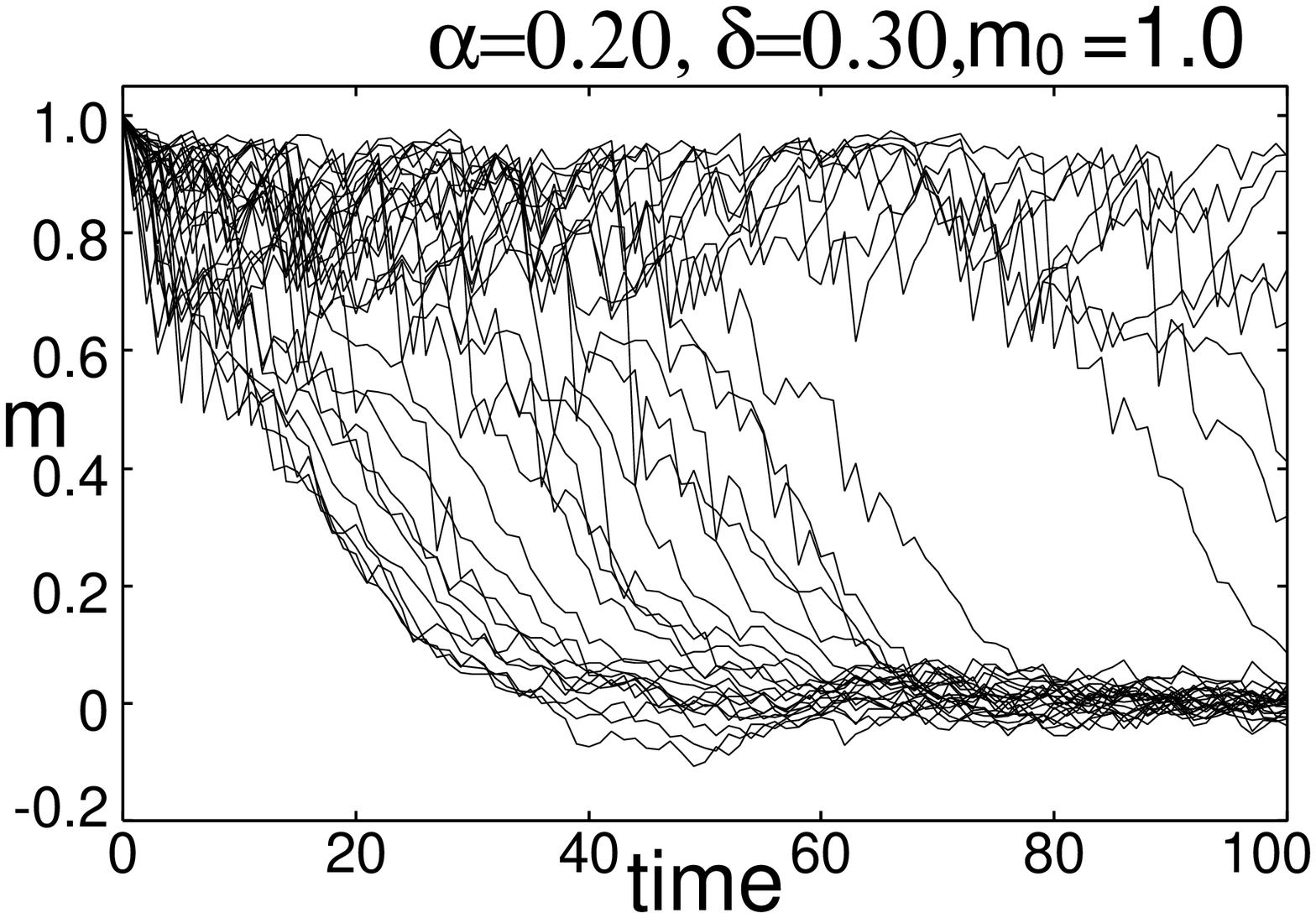} \\
  (c) $\delta=0.3$
 \end{center}
 \caption{Time evolutions of overlap with common synaptic inputs
 for (a) $\delta=0.1$, (b) $\delta=0.2$, and (c) $\delta=0.3$, where
 $\alpha=0.20$ and $m_0=1.0$.}
 \label{fig:ovlp}
\end{figure}

Next, we show the time evolutions of overlap for various $\delta$ values
in order to find the effect of the common synaptic
inputs. Figure~\ref{fig:ovlp} shows the overlaps for $\delta=0.1,0.2$
and $0.3$.  We use an initial overlap of $m_0=1$ in order to discuss the
stability of the memory state. For small $\delta$ values, as in
Figs.~\ref{fig:ovlp}(a) and \ref{fig:ovlp}(b), the stored patterns can
be stably retrieved. For $\delta=0.3$ as in Fig.~\ref{fig:ovlp}(c),
however, the network gradually reaches away from the memory state in many
of the samples.

\begin{figure}[tb]
 \begin{center}
  \includegraphics[width=75mm]{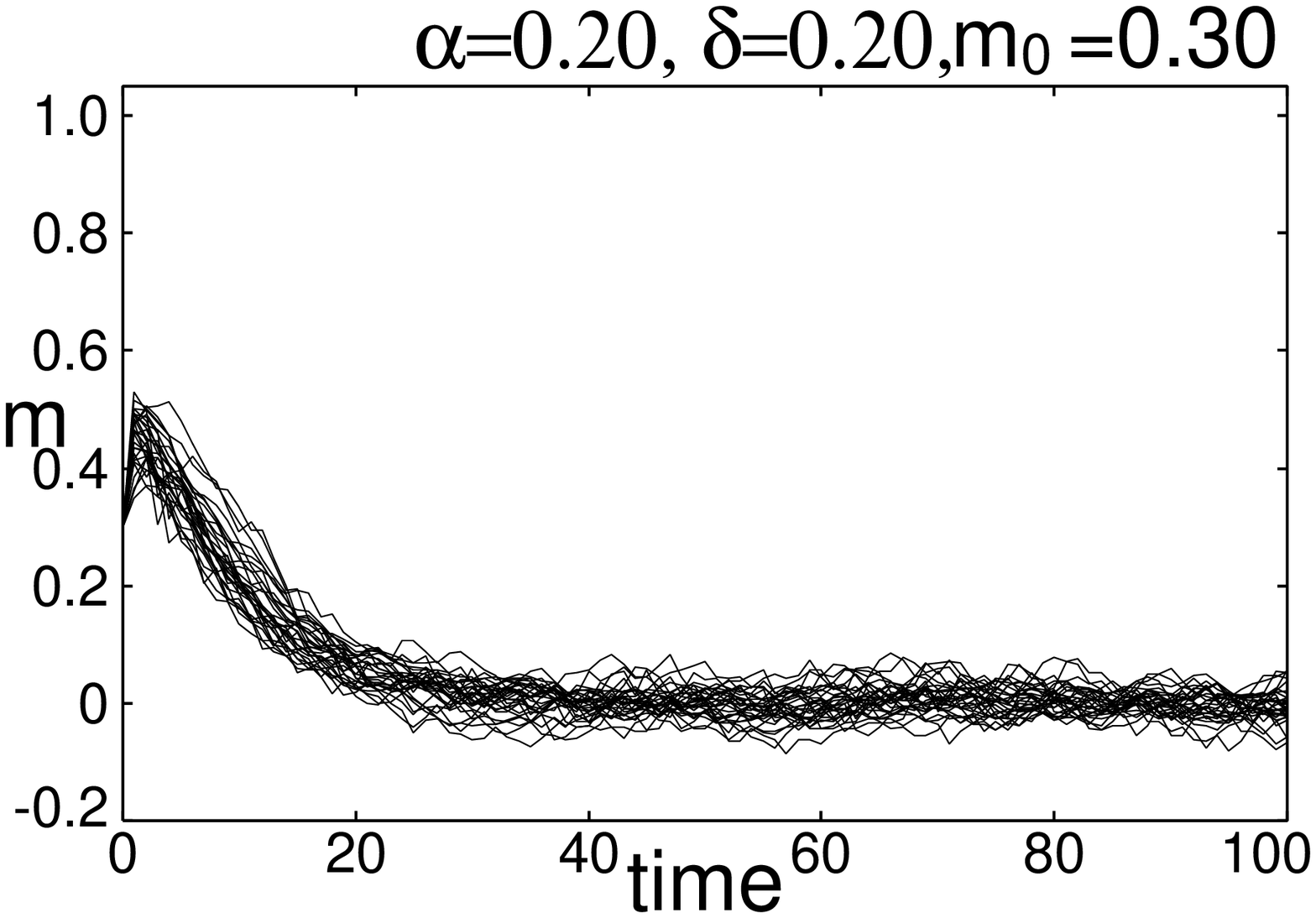} \\
  (a) $m_0=0.30$

  \includegraphics[width=75mm]{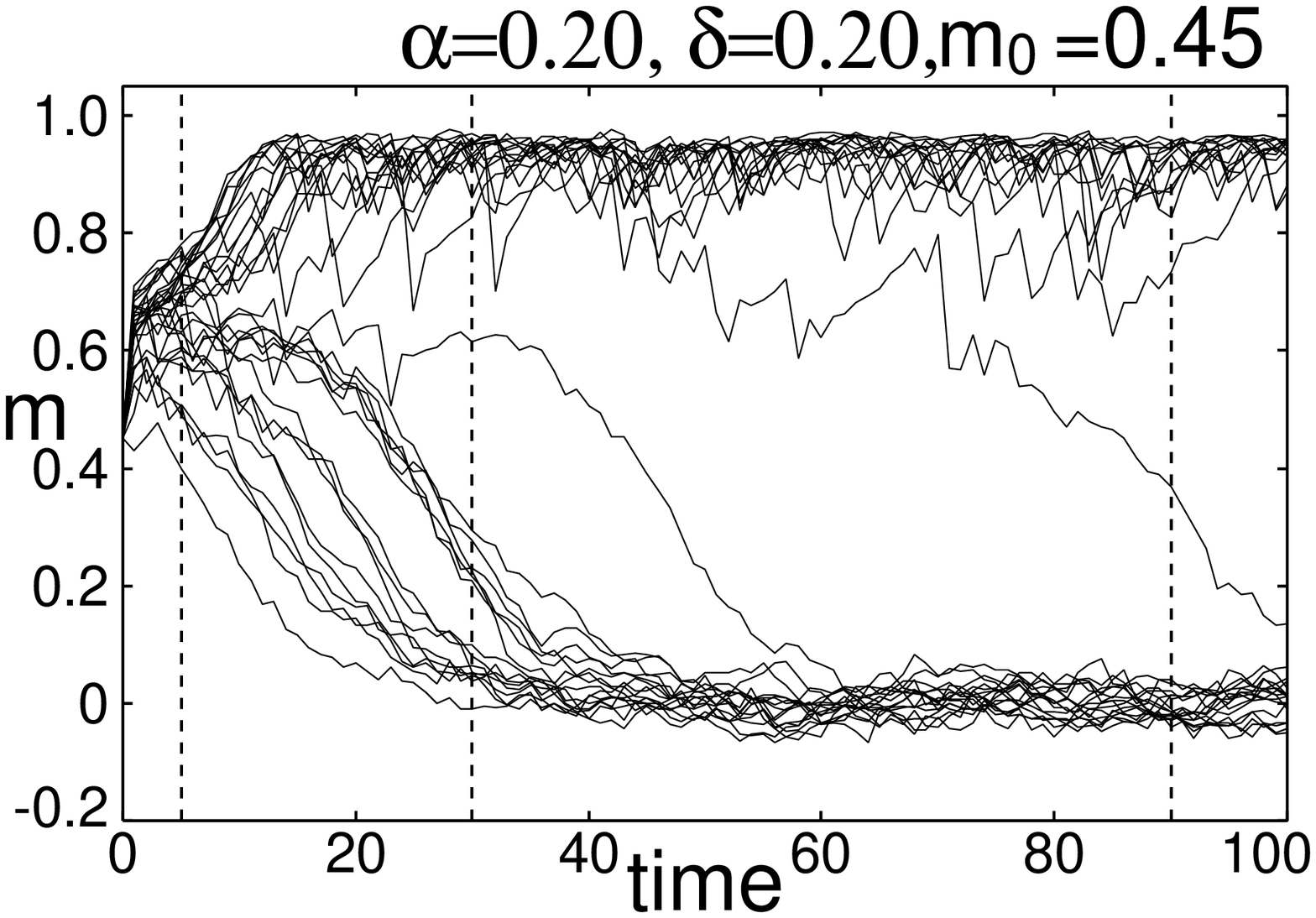} \\
  (b) $m_0=0.45$
 \end{center}
 \caption{Time evolutions of overlap for (a) $m_0=0.30$ and (b)
 $m_0=0.45$, where $\alpha=0.20$ and $\delta=0.20$.}
  \label{fig:overlap}
\end{figure}

Furthermore, we verify that a memory state is an attractor when the
memory state is stable. Figure~\ref{fig:overlap} shows the time evolutions
of overlap for $\delta=0.20$, where $m_0=0.30$ and $0.45$, $\alpha=0.20$ and
$N=5,000$. The figure shows $30$ samples of different trials. Whereas
the network reaches the nonretrieval state for all samples in the case
of $m_0=0.30$, as in Fig.~\ref{fig:overlap}(a), it reaches either the retrieval
or nonretrieval state depending on the samples in the case of $m_0=0.45$,
as in Fig.~\ref{fig:overlap}(b). From these results, the memory state is
the stationary state of the network, and it has a finite basin of
attraction. From the same initial state the network can reach different
attractors by the common synaptic inputs, which means that sample
dependence exists and that self-averaging breaks down
\cite{Amari_etal2003,YamanaOkada2005}.

\section{Probability Distribution}

From Figs.~\ref{fig:ovlp} and \ref{fig:overlap} we can see the sample
dependence by the common synaptic inputs. We therefore need to discuss
the distribution of macroscopic states instead of the behavior for each
trial in order to analyze the behavior of the network; that is, we must
discuss the probability distribution of overlaps. From
eq.~(\ref{eqn:PnextK}), the probability distribution at any time $t$ can
be evaluated when the initial distribution $p(m_0,\sigma_0)$ is
given. Here, we introduce the marginal probability distribution,
$p\left(m_t\right)$, which is integrated with respect to $\sigma_t$ :
\begin{equation}
 p\left(m_{t}\right)=\int d\sigma_t\; p\left(m_{t},\sigma_{t}\right).
  \label{eqn:Pmt}
\end{equation}

\begin{figure}[tb]
 \begin{center}
  \includegraphics[width=75mm]{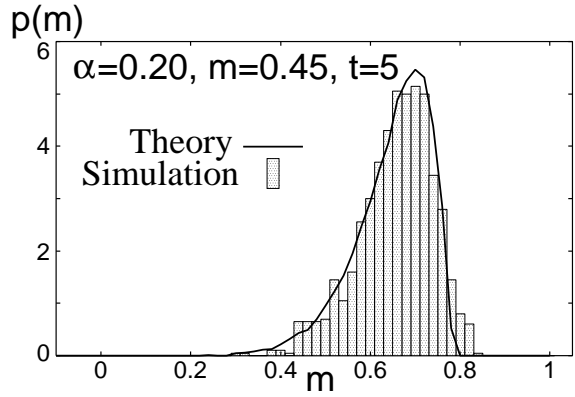} \\
  (a) $p(m_{5})$ at $t=5$
  
  \includegraphics[width=75mm]{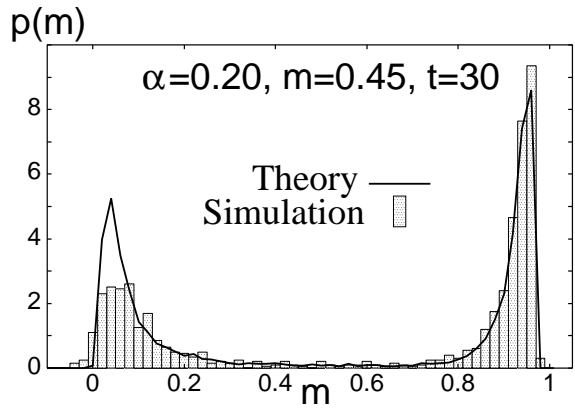}\\
  (b) $p(m_{30})$ at $t=30$

  \includegraphics[width=75mm]{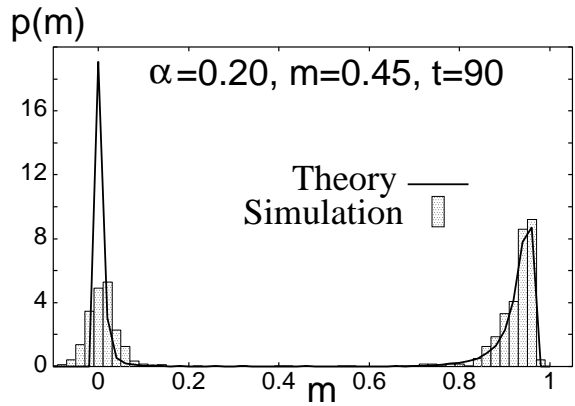} \\
  (c) $p(m_{90})$ at $t=90$
 \end{center}
 \caption{Marginal probability distribution at (a) $t=5$, (b) $t=30$,
 and (c) $t=90$, where $\alpha=0.20, m_0=0.45$ and $\delta=0.20$.}
 \label{fig:dist}
\end{figure}

\begin{figure}[tb]
 \begin{center}
  \includegraphics[width=75mm]{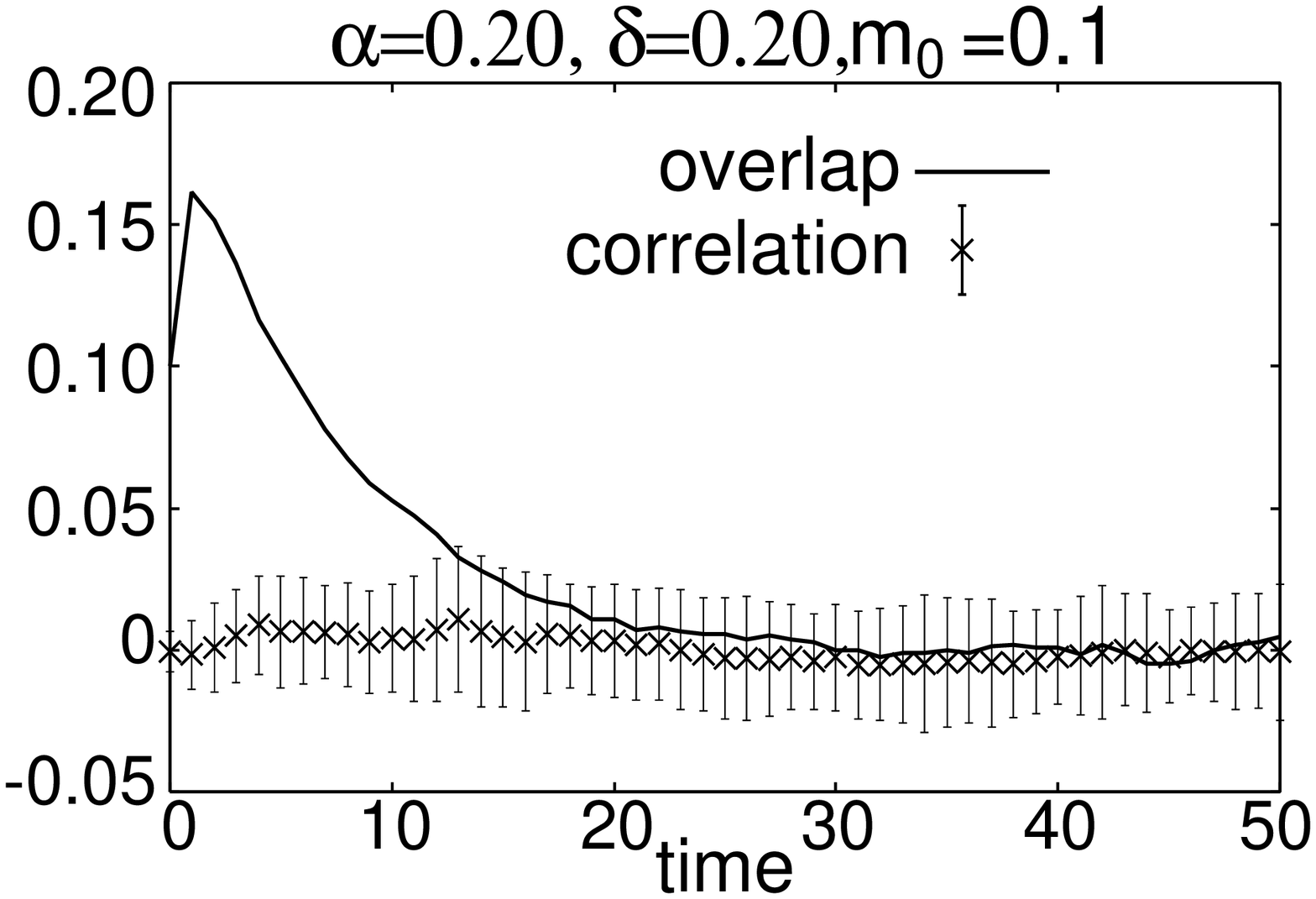}
  \caption{Time evolutions of overlap and time correlation coefficient,
  where $\alpha=0.20, m_0=0.10, \delta=0.20$.}
  \label{fig:correlation}
 \end{center}
\end{figure}

We analyze the probability distribution of overlaps at time $t=5, 30$
and $90$ in Fig.~\ref{fig:overlap}(b). Figure~\ref{fig:dist} shows the
marginal probability distribution obtained by our theory and histograms
obtained from the computer simulations. The lines denote the results
obtained from eq.~(\ref{eqn:Pmt}), and the boxes denote the histograms
for $1,000$ samples obtained from the computer simulations
($N=5,000$). In the cases of Figs.~\ref{fig:dist}(a) and
\ref{fig:dist}(b), the results obtained by the theory agree with those
obtained by the computer simulations.  On the other hand, in the case of
Fig.~\ref{fig:dist}(c), both results agree at $m_t\approx 1$, but the
distribution by the computer simulations spreads at $m_t\approx 0$. In
the nonretrieval state, the assumption of
$\vec{x}^t\approx\vec{\xi}^t$ may not be satisfied, in which case the
time correlation may not be ignored. We have, however, verified 
with the computer simulations that $\vec{x}^t$ has no time correlation
in the nonretrieval state.
Figure~\ref{fig:correlation} shows the time evolutions of overlap $m_t$
and the time correlation coefficient for an initial overlap $m_0=0.10$,
where $\alpha=0.20, \delta=0.20$ and $N=20,000$. The error bars and the
line represent the average and standard deviation of the time
correlation coefficients and the average of overlaps over 20 trials,
respectively.  The network state goes to the nonretrieval state, since
the overlap $m_t$ becomes zero. The time correlation coefficients are
calculated using the states $x_i^t$ and $x_i^{t+1}$. Since they are
almost zero, $\vec{x}^t$ has no time correlation.

Another possible source of disagreement in the nonretrieval case is in
the Gaussian assumption of the crosstalk noise. Although in the
sequential associative memory model without the common synaptic inputs
the crosstalk noise obeys the Gaussian distribution even in the
nonretrieval case \cite{Kawamura2002}, it is difficult to show that it
is the Gaussian in the model with the common synaptic inputs. Therefore,
we suppose that the fluctuation at $m_t\approx0$ is caused by either the
breakdown of the Gaussian assumption or the fact that there is a finite
number of neurons, as is the case with $\delta=0$
(Fig.~\ref{fig:delta0}).

\section{Summary}

Correlated firing such as that by synfire chains is a noticeable
phenomenon. The mechanism will be elucidated by theoretical models in
the future. We discussed the effects of the common synaptic inputs in a
sequential associative memory model.
In this model, correlated firing occurs because the input to each neuron
has a correlation due to the common synaptic inputs; therefore, sample
dependence exists.  We verified the existence of sample dependence via
computer simulations.
In order to investigate the correlated firing, we need to analyze
theoretically novel phenomena caused by the sample dependence. However,
we were unable to use the independence of units or neurons at the
thermodynamic limit. Moreover, in recurrent neural networks, theoretical
treatment is much more difficult because of feedback connections. We
therefore considered the sequential associative memory model, in which
time correlation can be ignored, allowing us to derive a recurrence
relation form of the PDF at the macroscopic state. The probability
distributions obtained by our theory agree with those obtained by the
computer simulations.

We analyzed the sequential associative memory model that had common
synaptic inputs. However, it may be hard to rigorously analyze models
such as autoassociative memory models since the time
correlation cannot be neglected.

\section*{Acknowledgments}

This work was partially supported by Grant-in-Aid for Scientific
Research on Priority Areas No. 14084212 and Grant-in-Aid for
Scientific Research (C) No. 14580438.


\end{document}